\documentclass[10pt,twocolumn,letterpaper]{article}

\usepackage{cvpr}
\usepackage{times}
\usepackage{epsfig}
\usepackage{graphicx}
\usepackage{amsmath}
\usepackage{amssymb}

\usepackage[pagebackref=true,breaklinks=true,letterpaper=true,colorlinks,bookmarks=false]{hyperref}
\usepackage{multirow}

\cvprfinalcopy 

\begin{document}

\title{A deep cascade of ensemble of dual domain networks with gradient-based T1 assistance and perceptual refinement for fast MRI reconstruction}

\author{Balamurali Murugesan\\
\and
Sriprabha Ramanarayanan\\
\and 
Sricharan Vijayarangan
\and 
Keerthi Ram
\and
Naranamangalam R Jagannathan
\and 
Mohanasankar Sivaprakasam
}

\maketitle

\begin{abstract}
Deep learning networks have shown promising results in fast magnetic resonance imaging (MRI) reconstruction. In our work, we develop deep networks to further improve the quantitative and the perceptual quality of reconstruction. To begin with, we propose reconsynergynet (RSN), a network that combines the complementary benefits of independently operating on both the image and the Fourier domain. For a single-coil acquisition, we introduce deep cascade RSN (DC-RSN), a cascade of RSN blocks interleaved with data fidelity (DF) units. Secondly, we improve the structure recovery of DC-RSN for T2 weighted Imaging (T2WI) through assistance of T1 weighted imaging (T1WI), a sequence with short acquisition time. T1 assistance is provided to DC-RSN through a gradient of log feature (GOLF) fusion. Furthermore, we propose perceptual refinement network (PRN) to refine the reconstructions for better visual information fidelity (VIF), a metric highly correlated to radiologist’s opinion on the image quality. Lastly, for multi-coil acquisition, we propose variable splitting RSN (VS-RSN), a deep cascade of blocks, each block containing RSN, multi-coil DF unit, and a weighted average module. We extensively validate our models DC-RSN and VS-RSN for single-coil and multi-coil acquisitions and report the state-of-the-art performance. We obtain a SSIM of 0.768, 0.923, 0.878 for knee single-coil-4x, multi-coil-4x, and multi-coil-8x in fastMRI. We also conduct experiments to demonstrate the efficacy of GOLF based T1 assistance and PRN. 
\end{abstract}

\section{Introduction}
\label{sec:introduction}
Magnetic resonance imaging (MRI) is a valuable diagnostic imaging modality that provides excellent spatial resolution with a superior soft-tissue contrast. However, MRI is an inherently slow acquisition process as the data sampling is carried out sequentially in k-space and the speed at which k-space can be traversed is limited by physiological and hardware constraints. These long acquisition times impose significant demands on patients, making the imaging modality expensive and less accessible \cite{hollingsworth}. Data acquisition can be accelerated by acquiring fewer k-space samples, which upon reconstruction provides a degraded image. Several works have been proposed to improve the reconstruction quality, including parallel imaging (PI) \cite{sense,grappa}, compressed sensing (CS) \cite{cs_mri} , and a combination of PI and CS \cite{cs_pi_1,cs_pi_2}. Recently, methods based on deep learning have shown promising results. However, the quantitative and the perceptual quality of these methods could be improved by the following: (1) effectively utilizing the image and k-space domain data; (2) exploiting the additional information from other sequences; (3) optimizing the network for a metric which highly correlates with the image quality scores of radiologists; and (4) availing the multi-coil data. In our work, we propose deep networks considering the above discussed possibilities.

\begin{figure*}
    \centering
    \includegraphics[width=0.9\linewidth]{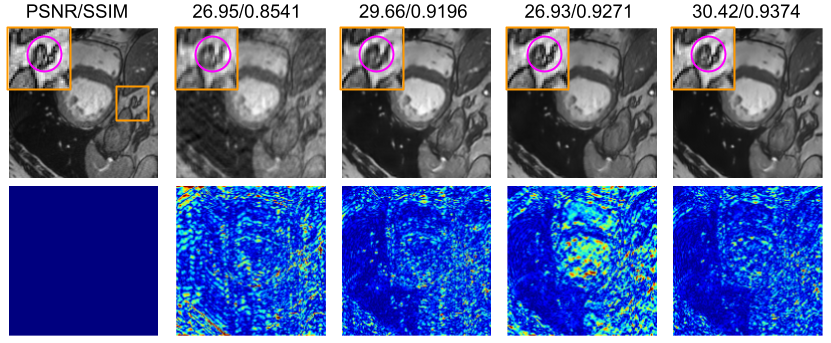}
    \caption{From left to right: Target, ZF, KI, II, and RSN. From top to bottom: Reconstructed image and its reconstruction error with respect to the target. Compared to II, KI has lower reconstruction error (higher PSNR). The bottom row shows that the residue of KI is lower compared to II. Compared to KI, the structure recovery of II is better (higher SSIM). The structures shown in the specific ROI are blurred and merged in KI while in II the structures are sharp and clearly separated. RSN provides both lower reconstruction error and better structure recovery (higher PSNR and SSIM).}
    \label{fig:rsn_illustration}
\end{figure*}

Firstly, the zero filled reconstruction (ZF) which is the inverse Fourier transform (IFT) of under sampled (US) k-space provides an image with aliasing artifacts. Deep learning networks have been developed to restore the original image by imputing the missing k-space \cite{automap} or by de-aliasing the degraded image \cite{wang}. Interestingly, we observed that networks operating on the Fourier domain provided lower reconstruction error while networks operating on the image domain provided a better structure recovery. The stacked version of the predictions from different models can effectively be combined through a fusion model \cite{stacked_generalization}. Hence, we propose reconsynergynet (RSN), a fusion (Fu) network which synchronously operates on the image domain outputs of both k-space to image (KI) and image to image (II) networks. KI network operates on the frequency domain while the II network operates on the image domain. A deep cascade convolutional neural network (DC-CNN) \cite{dc_cnn} incorporated the practical strategies of CS in deep learning through cascades of CNN interleaved with data fidelity (DF) units. CNN blocks are used for de-aliasing, while DF units are used to provide data consistency in the Fourier domain. Motivated by this adaptation, we propose DC-RSN, a deep cascade of RSN blocks interleaved with DF units for single-coil reconstruction. RSN fuses the de-aliased output of II network and the domain transformed output of KI network through Fu network, unlike the blocks in DC-CNN which only does de-aliasing. Fig. 1 demonstrates the effectiveness of operating in both the k-space and the image domain by comparing RSN with KI and II networks. 

Secondly, in routine radiological imaging, T1 weighted imaging (T1WI) and T2 weighted imaging (T2WI) are the two basic MR sequences used for diagnosis. T2WI is usually slower than T1WI due to the use of relatively longer repetition time (TR) and echo time (TE). Moreover, these sequences are structurally correlated, which facilitates the use of fully sampled (FS) T1WI for accelerating the acquisition of T2WI. Recently, the assistance of T1WI for T2WI reconstruction has been investigated \cite{t1_assistance}. We further explored this aspect and propose a gradient based T1 assistance to DC-RSN in two stages. In the first stage, we introduce DC-RSN-T1, a network similar to DC-RSN, except that for every cascade, Fu of RSN takes FS T1WI as an additional channel along with the reconstructions from KI and II networks to provide the structures missed by the KI and II networks. We encode the reconstruction from DC-RSN-T1 as a feature representation using gradient of log feature (GOLF) module which provides the feature corresponding to the logarithm of gradient of the input image. In the second stage, we fuse the intermediate feature maps of KI and II of RSN for every cascade of DC-RSN with the feature representation obtained from the first stage. We name this architecture as DC-RSN-T1-GOLF, where the GOLF obtained from the reconstruction of DC-RSN-T1 is used to guide the reconstruction of DC-RSN by explicitly providing the boundary information.

Thirdly, the deep learning networks developed for MRI reconstruction typically use the mean square error (MSE) as an objective function. However, MSE is often associated with over-smoothed edges and overlooks the subtle image textures critical for human perception. In computer vision, the perceptual quality of the image is improved by generative adversarial networks (GAN), which uses adversarial loss in addition to MSE \cite{esrgan_perceptual}. Perceptual index (PI), a no reference metric, is used in the vision community to validate the perceptual quality of an image \cite{perception_distortion_tradeoff}. Likewise, a recent study \cite{radiology_perception} reported that the metric visual information fidelity (VIF) \cite{vif} is highly correlated with the radiologist’s perception on the quality of MRI. In our work, we propose a CNN based perceptual refinement network (PRN) to refine the reconstructions of models trained using MSE for better perceptual quality. We also show that improvement in VIF can be obtained by training the PRN in an adversarial setup. PRN is successively connected to the pre-trained DC-RSN-T1-GOLF to form DC-RSN-T1-GOLF-PRN, an ensemble of dual domain cascade network with gradient-based T1 assistance and perceptual refinement for single-coil acquisition. 

Finally, multi-coil acquisition is the default option for many scan protocols and is supported by almost all modern clinical MRI scanners. Furthermore, for the same acceleration, reconstructions obtained from multi-coil acquisition are better and more tractable than the one from single-coil acquisition because of the information redundancy in multiple channels (Knoll et al., 2019). Variational network (VN) \cite{variational} and variable splitting network (VS-Net) \cite{vs_net} were proposed to specifically work for multi-coil acquisition. DF proposed in VS-Net for multi-coil acquisition is computationally efficient compared to DF in VN. Besides, DF in VS-Net is the direct extension of point wise data consistency operation in DC-CNN, unlike DF in VN, which is an approximate estimate through gradient descent. Inspired by VS-Net, we propose VS-RSN, a deep cascade of multi-coil specific blocks with each block containing RSN, DF unit, and weighted average module. RSN works on sensitivity-weighted multiple channels, DF unit does data consistency across multiple channels, and the weighted average module combines the reconstructions from RSN and DF. To the best of our knowledge, VS-RSN is the first dual domain cascade network for multi-coil reconstruction. Similar to DC-RSN, the image quality of VS-RSN are also improved through gradient assistance and perceptual refinement. 

In summary, the main contributions of our work are the following:
\begin{itemize}
    \item We propose novel dual domain cascade architectures DC-RSN and VS-RSN, for single-coil and multi-coil acquisition, respectively. 
    \item We propose GOLF based T1 assistance to provide more faithful reconstruction of T2WI.
    \item We propose PRN to refine the final reconstruction for obtaining high image quality scores from radiologists. 
    \item We conduct extensive comparison and show that our network DC-RSN for single-coil and VS-RSN for multi-coil are better than the respective state-of-the-art methods across acceleration factors and datasets.
    \item We conduct extensive experiments and demonstrate the efficacy of GOLF based T1 assistance in T2WI reconstruction. Furthermore, we extensively validate PRN with the proposed models and observe that PRN addition improves VIF.
    \item We validate DC-RSN and VS-RSN using the single and multi-coil knee dataset of fastMRI  (Zbontar et al., 2018). We obtain a competitive SSIM of 0.768, 0.923, 0.878 for knee single-coil-4x, multi-coil-4x, and multi-coil-8x, respectively.
\end{itemize}

The paper is organized as follows: Section 2 reviews the related works while section 3 provides the description of datasets used in the experiments. The design of DC-RSN, VS-RSN, GOLF based T1 assistance, and PRN are described in Section 4. Section 5 presents the results and discussions while section 6 contains the conclusions. 

\section{Related Work}
\label{sec:related_work}    
\subsection{Single-coil MRI reconstruction}
\label{subsec:related_work_single_coil}
\subsubsection{k-space to Image methods}
\label{subsubsec:related_work_single_coil_k-space_to_image}
These are networks that learn the mapping between US k-space and FS image. AUTOMAP \cite{automap} used fully connected (FC) network to learn the mapping between k-space and image domain. The major drawback with AUTOMAP is that the parameters of the network increases quadratically with increase in input k-space dimension. This makes the usage of AUTOMAP difficult for k-space with higher dimensions (like 256x256). To overcome this limitation, dAUTOMAP \cite{dautomap} replaced the FC layers in AUTOMAP by a fully convolutional network using the separability property of 2D IFT. 

\subsubsection{Image to Image methods}
\label{subsubsec:related_work_single_coil_image_to_image}
These are networks that learn the mapping between US and FS image. Simple convolutional networks \cite{wang} have been used to learn the mapping, and it has been shown that learning the aliasing artifact is efficient compared to learning alias-free FS image \cite{residual_conference}. RefineGAN (ReGAN) \cite{cyclic_loss} and DAGAN \cite{dagan} used GAN framework with UNet (Ronneberger et al., 2015) like network as a generator and classic deep learning classifier as a discriminator. Both these networks used linear combination of adversarial loss, image domain loss, and frequency domain loss as their objective function. DAGAN also tried to improve the perceptual quality with additional VGG loss.

\subsubsection{Cascade methods}
\label{subsubsec:related_work_single_coil_cascade}
Cascade networks help to learn the mapping between US and FS image through unrolled optimization of image to image learning and data consistency in Fourier domain. DC-CNN \cite{dc_cnn} proposed to utilize cascades of CNN for image reconstruction while DF layers were used for data consistency. DC-UNet \cite{dc_unet} replaced CNN in DC-CNN with UNet. Likewise, DC-RDN \cite{recursive_dilated} used a recursive dilated network in place of CNN in DC-CNN. In DC-DEN \cite{dc-ensemble}, the features extracted from each CNN block were connected to other CNN blocks through dense connections which were subsequently concatenated to obtain the final reconstruction.

\subsubsection{Hybrid methods}
\label{subsubsec:related_work_single_coil_hybrid}
These are networks that operate on both the k-space and the image domain apart from k-space data consistency operations. KIKI-Net \cite{kikinet} proposed cascade of k-space and image CNN interleaved by DF units. K-space CNN was used to obtain FS k-space from US k-space. Image CNN was used to obtain FS image from the IFT of the predicted FS k-space. DC-Hybrid \cite{hybrid} used a similar architecture as that of KIKI-Net to operate on both the k-space and the image domains. However, the DC-Hybrid architecture started with image domain operation and followed it with the k-space domain operation.

\subsection{Multi-coil MRI reconstruction}
\label{subsec:related_work_multi_coil}
Similar to DC-CNN, the architectures developed for multi-coil acquisition mimic the classic iterative image reconstruction. VN \cite{variational} proposed to utilize cascades of image CNN interleaved by DF through gradient descent scheme. MoDL \cite{modl} proposed to use cascades of CNN whose parameters are shared, thereby reducing the parameter complexity. Unlike VN, MoDL used conjugate-gradient setup for DF. VS-Net \cite{vs_net} proposed to replace the gradient and the conjugate-gradient update of VN and MoDL for DF with a point-wise analytical solution, making VS-Net computationally efficient and numerically accurate. 

\section{Dataset Description}
\subsection{Single-coil dataset}
\subsubsection{Real-valued MRI data}
\textbf{Cardiac dataset} \cite{cardiac_dataset}: Automated cardiac diagnosis challenge (ACDC) consists of 150 and 50 patient records for training and validation, respectively. We extract the 2D slices and crop it to 150 x 150, which amounted to 1841 and 1076 slices for training and validation, respectively.

\textbf{Kirby dataset} \cite{kirby_dataset}: The human brain dataset consists of 42 T1 MPRAGE volumes with dimensions 256 x 256. We consider the center 90 slices from each volume, which gave 29 volumes with 2610 slices for training and 13 volumes with 1170 slices for validation. 

\subsubsection{Complex-valued MRI data}
\textbf{Calgary dataset} \cite{calgary_dataset}: The human brain dataset has 45 T1 volumes with dimensions 256 x 256. The data was acquired with 12-channel receiver coil which was combined to simulate a single-coil acquisition. We consider the center 110 slices from each volume, which provided 25 volumes with 2750 slices and 10 volumes with 1100 slices for training and validation, respectively. 

\subsection{T1-T2 paired dataset}
\textbf{MRBrainS} \cite{mrbrains_dataset} dataset contains paired T1 and T2-FLAIR volumes of 7 subjects. The dimensions of the volumes are 240 x 240. We use T1 volumes to assist T2-FLAIR reconstruction. We utilize the data from 5 subjects with 240 slices for training and 2 subjects with 96 slices for validation. 

\subsection{Multi-coil dataset}
\textbf{Knee dataset} \cite{variational}: The dataset has five image acquisition protocols: coronal proton-density (PD), coronal fat-saturated PD, axial fat-saturated T2, sagittal fat-saturated T2 and sagittal PD. The data was acquired using 15 channel receiver coil for 20 patients. Each patient data has 40 slices with 15 channels including their respective sensitivity maps. We consider the center 20 slices for the experiments. We split the patient data into 10 with 200 slices for training and remaining 10 with 200 slices for validation. 

\begin{figure*}
    \centering
    \includegraphics[width=0.9\linewidth]{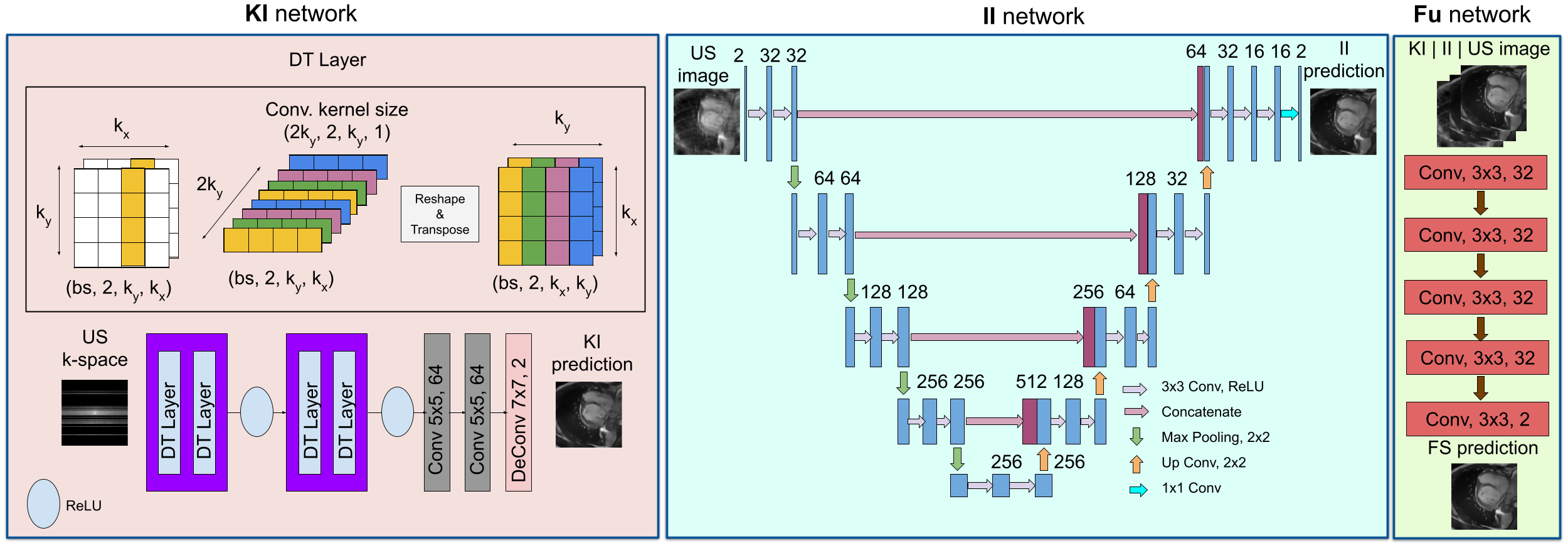}
    \caption{Outline of k-space to image (KI), image to image (II) and fusion (Fu) networks in ReconSynergyNet (RSN)}
    \label{fig:rsn}
\end{figure*}

\begin{figure*}
    \centering
    \includegraphics[width=0.9\linewidth]{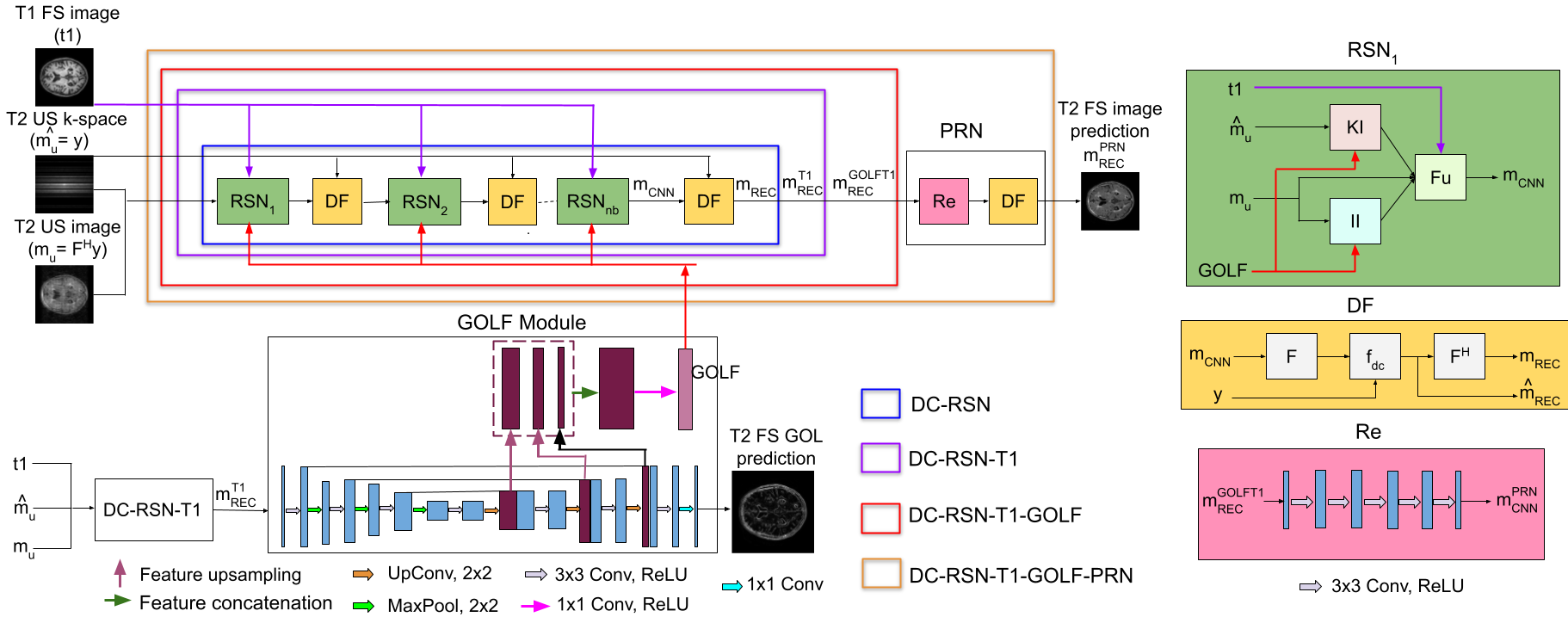}
    \caption{Outline of Deep Cascade RSN (DC-RSN) with Gradient of Log Feature (GOLF), T1 assistance and Perceptual Refinement Network (PRN).}
    \label{fig:dc_rsn}
\end{figure*}

\begin{figure*}
    \centering
    \includegraphics[width=0.9\linewidth]{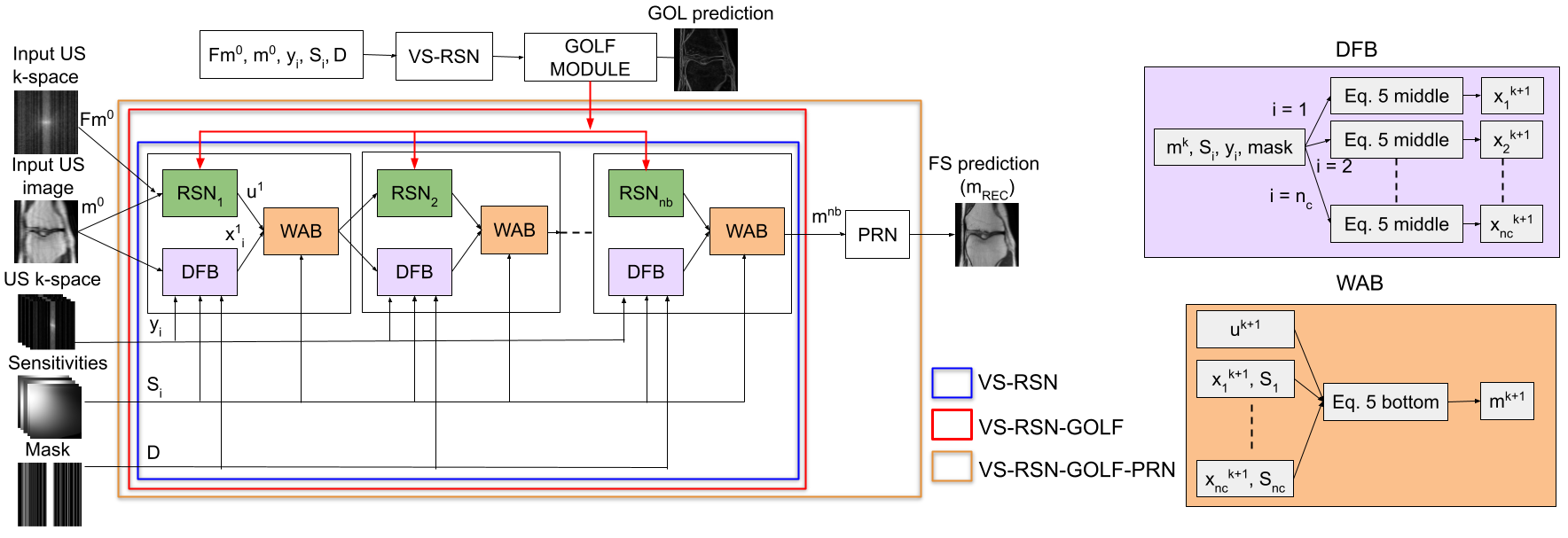}
    \caption{Outline of Variable splitting ReconSynergyNet (VS-RSN) with Gradient of Log Feature (GOLF) and Perceptual Refinement Network (PRN).}   
    \label{fig:vs_rsn}
\end{figure*}

\section{Methodology}
\label{sec:methodology}

\subsection{Problem formulation}
\label{subsec:methodology_problem_formulation}

Let $m \in C^{N}$ be a column stacked vector of complex-value MR image and $y_{i} \in C^{M}$ ($M << N$) be the under-sampled k-space data with respect to the $i^{th}$ MR receiver coil. Recovery of $m$ from $y_{i}$ is an ill-posed inverse problem. According to compressed sensing (CS) theory, $m$ can be obtained by minimizing the following optimization problem:

\begin{equation}
    \label{eq:problem_formulation}
    \footnotesize
    \underset{m}{min}\quad\frac{\lambda}{2}\sum_{i=1}^{n_{c}}||DFS_{i}m - y_{i}||_{2}^{2} + R(m)
\end{equation}
where ${R(m)}$ is the sparse regularization term, $\lambda$ is the weight to balance the two terms, $D \in R^{M \times N}$ is the undersampling matrix, $F \in C^{N \times N}$ is the Fourier transform matrix, $n_{c}$ denotes number of receiver coils and $S_{i} \in C^{N \times N}$ is the sensitivity map of $i^{th}$ coil. 

\subsection{Fundamental CNN block - RSN}
\label{subsec:methodology_fundamental_cnn_block}    
We propose RSN, a basic block for MRI reconstruction which consists of the following components (Fig. \ref{fig:rsn}):

\textbf{KI network:}
We consider dAUTOMAP as KI network instead of the common k-space CNN followed by IFT as it offers equivalent performance of state-of-the-art AUTOMAP with linear complexity. Let $p$ be the vector form of image $P$, $q$ be the vector form of k-space $Q$. Then, by the theory of Fourier transform: 
\begin{equation}
    \label{eq:dautomap_formulation}
    \footnotesize
    p = F^{*}q = (F_{1}^{*} \otimes F_{2}^{*})q  = vec(F_{1}^{*}QF_{2}^{*}) = vec((F_{2}^{*}(F_{1}^{*}Q)^T)^T)
\end{equation}
where $F^{*}$ is conjugate of 2D Discrete Fourier transform (DFT) $F$ and $F_{1}^{*}, F_{2}^{*}$ are conjugates of 1D DFT matrices. Note that $F_{1}^{*}Q$ can be realized by 1D convolution, this is termed as decomposed transform layer (DT layer). Hence, 2D DFT can be obtained by repeating DT and transpose operation twice. The predicted image is further refined by 2D convolutions. 

\textbf{II network:}
The II network is U-Net \cite{unet}, a popular multiscale network for structure recovery. U-Net is a encoder-decoder architecture which uses convolutions (for extracting features), ReLU activations (to add non-linearity), max-pooling (downsampling) layers, up-convolution (upsampling) layers and skip connections (transfer features). 

\textbf{Fu network:}
An efficient fusion of the reconstructions of KI and II can help us provide an improved reconstruction. Wolpert \cite{stacked_generalization} showed that stacked version of the predictions from different models can be effectively combined through a fusion model. Interestingly, several works have used CNN for fusion as it's hard for a non-learning-based algorithm to combine the benefits of different models \cite{fusion_gland}\cite{fusion_hyperdensenets}. Similarly, SRCNN \cite{srcnn} showed that by stacking channels with high cross-correlation at the input, the convolution layers can leverage the natural correspondences between the channels for better reconstruction. In our case the correlated channels are reconstructions from KI and II networks. The Fu network consists of five 3x3 convolution layers with ReLU activations. We also considered stacking the input image (II network's input) along with the outputs of KI and II networks to provide an idea of unsmoothed structures to Fu network.

\subsection{DC-RSN: Single-coil MRI reconstruction}
\label{subsec:methodology_single_coil_mri}
For single-coil MRI reconstruction, Eq. \ref{eq:problem_formulation} is converted to the following:
\begin{equation}
    \label{eq:problem_formulation_dc_rsn_1}
    \footnotesize
    \underset{m,\theta}{min}\quad\frac{\lambda}{2}||DFm - y||_{2}^{2} + || m - f_{cnn}(m_{u}\lvert\theta) ||_{2}^2 
\end{equation}
where $f_{cnn}$ is the deep network parameterised by $\theta$, which learns the mapping between $m_{u}$(undersampled column stacked complex-value MR image) and  $m$. To provide consistency with the acquired k-space data, the following data fidelity procedure is followed:
\begin{equation}
\label{eq:problem_formulation_dc_rsn_2}
\footnotesize
\hat{m}_{rec}=
\begin{cases}
  \hat{m}_{cnn}(k)  & \ k\notin\Omega \\
  \frac{\hat{m}_{cnn}(k) + \lambda \hat{m}_{u}(k)}{1+\lambda} & k\in\Omega \\
\end{cases}
\end{equation}
where $\hat{m}_{cnn} = Fm_{cnn}, m_{cnn}=f_{cnn}(m_{u}\lvert\theta)$, $\hat{m}_{u} = Fm_{u}$, $m_{rec}= F^{H}\hat{m}_{rec}$, $\Omega$ is an index set indicating which k space measurements have been sampled.

We propose DC-RSN (Fig. \ref{fig:dc_rsn}) for single-coil MRI reconstruction. DC-RSN consists of $n_{b}$ cascades of RSN ($f_{cnn}$) blocks and DF layers. RSN takes in US k-space and image to provide predicted FS image, while DF takes the predicted FS image and returns data (image, k-space) which are consistent in Fourier domain.

\subsection{VS-RSN: Multi-coil MRI reconstruction}
\label{subsec:methodology_multi_coil_mri}
In order to optimize Eq. \ref{eq:problem_formulation} efficiently, VS-Net \cite{vs_net} developed a variable splitting method by introducing auxiliary splitting variables $u \in C^{N}$ and $\{x_{i} \in C^{N}\}^{n_{c}}_{i=1}$ and derived the final solution as given below:

\begin{equation} 
\label{eq:problem_formulation_vs_rsn}
\footnotesize
\begin{split}
u^{k+1} &= denoiser(m^{k})  \\ 
x_{i}^{k+1} &= F^{-1}((\lambda D^{T}D + \alpha I)^{-1}(\alpha FS_{i}m^{k} + \lambda D^{T}y_{i})) \\ 
m^{k+1} &= (\beta I + \alpha \sum_{i=1}^{n_{c}}S_{i}^{H}S_{i})^{-1}(\beta u^{k+1} + \alpha \sum_{i=1}^{n_{c}}S_{i}^{H}x_{i}^{k+1})
\end{split}
\end{equation}

From the above equations, the following can be inferred: 1) Top equation: The original problem (Eq. \ref{eq:problem_formulation}) is converted to denoising problem. 2) Middle Equation: Provides data consistency to k-space for each coil. 3) Bottom equation: Computes a weighted average of the results obtained from the first two equations.

We propose VS-RSN (Variable Splitting - RSN) (Fig. \ref{fig:vs_rsn}) for multi-coil MRI reconstruction which can accommodate the iterative setup formulated in Eq. \ref{eq:problem_formulation_vs_rsn}. VS-RSN consists of $n_{b}$ cascades of three blocks: RSN as denoiser block, data fidelity block (DFB) and weighted average block (WAB). RSN takes in sensitivity-weighted US image ($m^{0} =  \sum_{i}^{n_{c}}S_{i}^{H}F^{-1}D^{T}y_{i}$) and it's respective k-space ($Fm^{0}$) as input. DFB uses pre-computed coil sensitivity maps ($\{S_{i}\}_{i=1}^{n_{c}}$), binary sampling mask ($D^{T}D$) and undersampled k-space data ($\{D^{T}y_{i}\}^{n_{c}}_{i=1}$) to provide data consistency in k-space for every coil. WAB uses coil sensitivity to weight the output of DFB and combine it with the output of RSN. Instead of pre-computing, the sensitivity maps could also be jointly learned with reconstruction using Auto-Calibration Signal of k-space as done in \cite{fb_varnet}.

\subsection{Assistance to MRI reconstruction}
\label{subsec:methodology_assistance_to_mri}

\subsubsection{Gradient assistance}
\label{subsubsec:methodology_gradient_assistance_to_mri}
Image gradients in the log-transformed domain can be used to guide image restoration tasks through its explicit boundary information \cite{riemannian_loss}. We call this type of gradient as GOL (Gradient of Logarithm). In our work, we propose to provide assistance to DC-RSN and VS-RSN through GOL of image. We build a GOLF (GOL feature) module with UNet and a single convolution layer. UNet architecture is trained on FS image and its GOL, while the features are extracted from the multiple levels of UNet, resized and concatenated to form a deep feature map which is then given to the single convolution layer with ReLU activation to provide the depth reduced effective feature map. These features are provided to RSN in each cascade of the DC-RSN and VS-RSN. In each RSN, the features are concatenated with the feature maps of 2D convolution and decoder layers of KI and II respectively. This design choice is made empirically. Feature concatenation explicitly provides the structural information required for reconstruction. We call the DC-RSN and VS-RSN with GOLF assistance as DC-RSN-GOLF and VS-RSN-GOLF respectively. During test time, the output of pretrained DC-RSN or VS-RSN is provided as input to the GOLF module for required features. The schematic of DC-RSN-GOLF and VS-RSN-GOLF can be found in Fig. \ref{fig:dc_rsn} and \ref{fig:vs_rsn} respectively. 

\subsubsection{T1 assistance}
\label{subsubsec:methodology_t1_assistance_to_mri}
The structural information in T1WI is highly correlated with T2WI. Hence, FS T1WI can be used to compensate for missing structures in US T2WI. In our work, we propose DC-RSN-T1 in which FS T1WI is provided as assistance to RSN at each cascade in DC-RSN. Specifically, the input to the Fu network is the channel stacked FS T1WI, KI and II network outputs. This design enables Fu network to effectively fuse FS T1WI with reconstructions obtained from KI and II networks. The notion behind this design is that FS T1WI could provide the structures which both KI and II networks would have failed to reconstruct because of missing frequencies and structures in k-space and image respectively. Fig. \ref{fig:dc_rsn} provides the outline. 

\subsubsection{Combined assistance - Gradient and T1}
\label{subsubsec:methodology_gradient_t1_assistance_to_mri}
We also propose DC-RSN-T1-GOLF with an optimal combination of GOLF and T1 assistance that maximizes the benefits of both. During test time of DC-RSN-T1-GOLF, instead of providing the output of DC-RSN to GOLF module, we provide the output of DC-RSN-T1. GOLF obtained using DC-RSN-T1 will have a feature representation closer to FS image compared to DC-RSN as DC-RSN-T1 would have reconstructed structures missed by DC-RSN. The enhanced GOLF is concatenated with the intermediate feature maps of KI and II of every cascade which provides the KI and II with explicit improved structural information. Fig. \ref{fig:dc_rsn} provides the illustration of DC-RSN-T1-GOLF.

\subsection{PRN for MRI reconstruction}
The prediction of pre-trained reconstruction networks are refined for better perceptual quality using PRN. PRN is a five layer CNN (Re network) followed by DF and is adversarially trained using WGAN \cite{wgan}. We use basic CNN as a refinement block so as to show its ability to improve the perceptual quality of the reconstructed image. DF is a necessary component for MRI reconstruction as it provides the required consistency in k-space. We choose WGAN for adversarial training as it provides more stability, better convergence and accurate estimate of the divergence between generator and data distributions. We use both adversarial and distortion loss for WGAN training. We provide higher weightage to adversarial loss compared to distortion loss as the adversarial component helps in providing perceptually better images \cite{perception_distortion_tradeoff}. We use PRN to refine the outputs of DC-RSN, VS-RSN and DC-RSN-T1-GOLF for better image perception. The overview can be found in Fig \ref{fig:dc_rsn} and \ref{fig:vs_rsn}. 

\section{Results and discussions}
\label{sec:experiments_and_results}

\subsection{Implementation details}
US data is retrospectively obtained using fixed cartesian undersampling masks for 4x and 5x acceleration. The source code of DC-CNN is used for its implementation. In the case of DC-UNet, CNN in DC-CNN is replaced with UNet from the fastMRI repository. The dense connections are added to CNN in DC-CNN to replicate the design of DC-DEN. Likewise, dilated CNNs with recursive connections are used in place of CNN in DC-CNN for DC-RDN. The alternative CNNs in DC-CNN are used to operate on image and k-space through their respective intermediate Fourier operations as demonstrated in the codebase of DC-Hybrid. UNet from fastMRI repository is used as the generator for DAGAN and ReGAN, while the designs of discriminator and loss functions were adapted from their respective repositories . The implementation of DenseUNet-T1 is taken from the publicly available code on semantic segmentation. VS-Net is implemented using its original repository. In the case of VN, the point based DF in VS-Net is replaced with the gradient descent based DF. From literature \cite{vs_net}, it is known that higher the number of cascades, better the reconstruction quality. Experiments demonstrating the same can be found in Figure A1 and Figure A2 of the supplementary material. In this work, due to resource constraints, the number of cascades is set to five for cardiac dataset and three for the remaining datasets.  Models are trained using MSE loss with Adam optimizer (Kingma and Ba, 2014). Adversarial models are trained using the combination of MSE and Wasserstein distance with Adam and SGD optimizers. DC-RSN, VS-RSN, and DC-RSN-T1 involve single stage training. Models with GOLF assistance (DC-RSN-GOLF, VS-RSN-GOLF, and DC-RSN-T1-GOLF) require two stage training. In the first stage, the base model (DC-RSN, VS-RSN, and DC-RSN-T1) is trained while in the second stage, training is done for the base model whose intermediate features are concatenated with the GOLF of first stage reconstruction. PRN is adversarially trained with inputs being the reconstructions of pre-trained networks (DC-RSN, VS-RSN and DC-RSN-T1-GOLF). PSNR and SSIM metrics are used to evaluate the reconstruction quality. VIF is used to validate the reconstruction for radiologist's opinion on image quality.

\subsection{Single-coil MRI reconstruction}
\subsubsection{Real-valued MRI data}

\begin{table*}
\scriptsize
\centering
\caption{Quantitative comparison of single-coil real valued MRI reconstruction architectures}
\label{tab:cardiac_kirby}
\begin{tabular}{|l|l|l|l|l|l|l|l|l|}
\hline
\multirow{3}{*}{Method} & \multicolumn{4}{c|}{Cardiac} & \multicolumn{4}{c|}{Kirby} \\ \cline{2-9} 
 & \multicolumn{2}{c|}{4x} & \multicolumn{2}{c|}{5x} & \multicolumn{2}{c|}{4x} & \multicolumn{2}{c|}{5x} \\ \cline{2-9} 
 & \multicolumn{1}{c|}{PSNR} & \multicolumn{1}{c|}{SSIM} & \multicolumn{1}{c|}{PSNR} & \multicolumn{1}{c|}{SSIM} & \multicolumn{1}{c|}{PSNR} & \multicolumn{1}{c|}{SSIM} & \multicolumn{1}{c|}{PSNR} & \multicolumn{1}{c|}{SSIM} \\ \hline
US & 24.27 $\pm$ 3.10 & 0.6996 $\pm$ 0.08 & 23.82 $\pm$ 3.11 & 0.6742 $\pm$ 0.08 & 25.7 $\pm$ 1.46 & 0.5965 $\pm$ 0.08 & 25.36 $\pm$ 1.47 & 0.5794 $\pm$ 0.08 \\ \hline
DAGAN\cite{dagan} & 28.52 $\pm$ 2.71 & 0.841 $\pm$ 0.04 & 28.02 $\pm$ 2.80 & 0.8248 $\pm$ 0.05 & 31.58 $\pm$ 1.30 & 0.8845 $\pm$ 0.01 & 30.93 $\pm$ 1.29 & 0.8719 $\pm$ 0.02 \\ \hline
DC-CNN\cite{dc_cnn} & 32.75 $\pm$ 3.28 & 0.9195 $\pm$ 0.04 & 31.75 $\pm$ 3.40 & 0.9054 $\pm$ 0.04 & 34.67 $\pm$ 1.78 & 0.9522 $\pm$ 0.01 & 33.31 $\pm$ 1.69 & 0.9415 $\pm$ 0.01 \\ \hline
DC-DEN\cite{dc-ensemble} & 33.22 $\pm$ 3.46 & 0.9249 $\pm$ 0.04 & 32.3 $\pm$ 3.57 & 0.9126 $\pm$ 0.04 & 35.27 $\pm$ 1.83 & 0.955 $\pm$ 0.01 & 33.73 $\pm$ 1.70 & 0.9425 $\pm$ 0.01 \\ \hline
DC-RDN\cite{recursive_dilated} & 32.95 $\pm$ 3.40 & 0.9233 $\pm$ 0.04 & 32.09 $\pm$ 3.57 & 0.9115 $\pm$ 0.04 & 35.61 $\pm$ 1.84 & 0.9629 $\pm$ 0.01 & 33.95 $\pm$ 1.70 & 0.95 $\pm$ 0.01 \\ \hline
DC-UNet\cite{dc_unet} & 33.17 $\pm$ 3.60 & 0.9276 $\pm$ 0.04 & 32.55 $\pm$ 3.71 & 0.9189 $\pm$ 0.04 & 36.4 $\pm$ 1.80 & 0.9697 $\pm$ 0.01 & 34.76 $\pm$ 1.67 & 0.9586 $\pm$ 0.01 \\ \hline
DC-RSN(Ours)& \textbf{33.61 $\pm$ 3.57} & \textbf{0.9322 $\pm$ 0.04} & \textbf{32.65 $\pm$ 3.67} & \textbf{0.92 $\pm$ 0.04} & \textbf{36.83 $\pm$ 2.0} & \textbf{0.9707 $\pm$ 0.01} & \textbf{35.22 $\pm$ 1.90} & \textbf{0.9609 $\pm$ 0.01} \\ \hline
\end{tabular}
\end{table*}

\begin{figure*}
    \centering
    \includegraphics[width=0.9\linewidth]{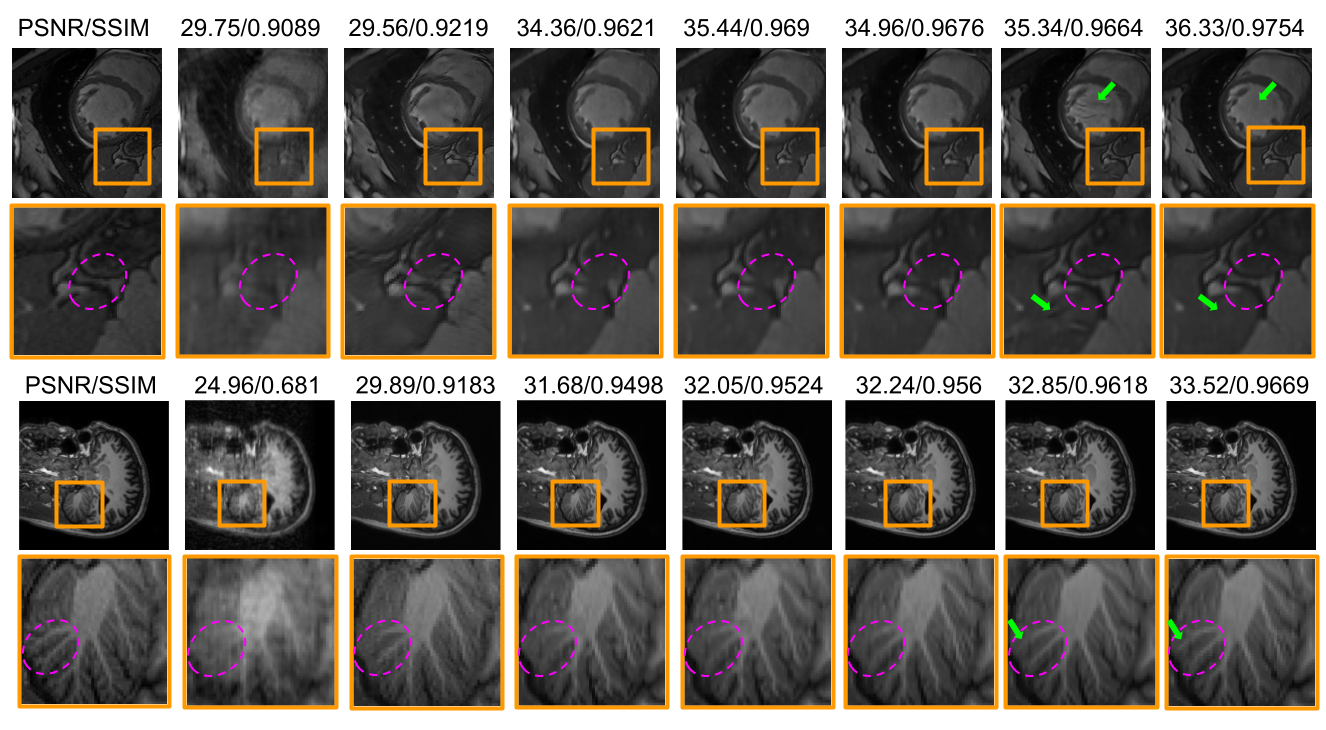}
    \caption{Qualitative comparison of single-coil real-valued MRI reconstruction architectures. From left to right (Target, ZF, DAGAN, DC-CNN, DC-DEN, DC-RDN, DC-UNet, and DC-RSN). Top row (Cardiac dataset): DAGAN has recovered the structure (pink dotted circle), but the reconstruction suffers from severe artifacts; DC-CNN, DC-DEN, and DC-RDN have completely missed the structure; DC-Unet and DC-RSN have properly recovered the structure, but DC-UNet has not removed the aliasing artifacts (green arrow). Bottom row (Kirby dataset): DAGAN has partially recovered the couple of structures in the region of interest (pink dotted circle); DC-CNN, DC-DEN, and DC-RDN have difficulty in delineating the structures; DC-Unet and DC-RSN have sharply recovered the structures, but DC-UNet has failed to recover the complete structure (green arrow).}
    \label{fig:cardiac_kirby}
\end{figure*}

In this experiment, cardiac and kirby datasets were used to compare our architecture DC-RSN with the architectures proposed for real-valued single-coil MRI reconstruction. The quantitative comparison of the architectures is presented in Table \ref{tab:cardiac_kirby}. It is clear from the table that DC-RSN fairs significantly better than other architectures in terms of PSNR and SSIM across different datasets and acceleration factors. This can be attributed to the RSN block which uses Fu network to effectively combine the benefits of simultaneously operating on both the k-space and the image domains. The qualitative comparison of the architectures for the datasets is depicted in Fig. \ref{fig:cardiac_kirby}, which shows that DC-RSN could reconstruct most intricate structures with reduced artifacts.

\subsubsection{Complex-valued MRI data}
\begin{figure*}
    \centering
    \includegraphics[width=0.9\linewidth]{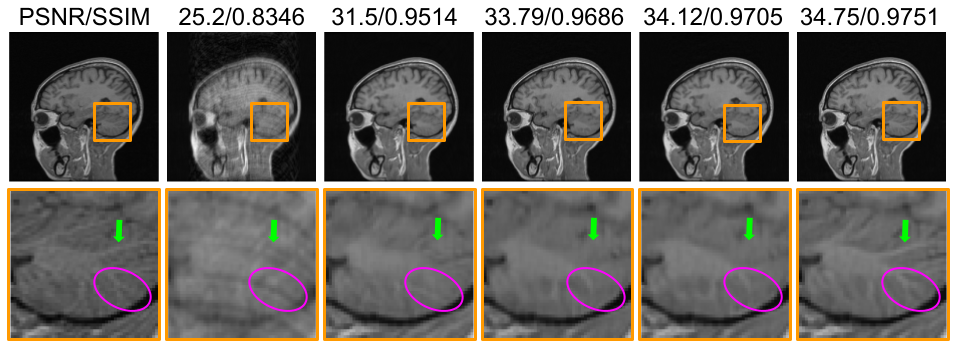}
    \caption{Qualitative comparison of single-coil complex-valued MRI reconstruction architectures. From left to right (Target, ZF, ReGAN, DC-CNN, DC-Hybrid, and DC-RSN). The structure denoted by green arrow in target is only recovered by DC-RSN while other architectures have failed to recover it. The structures in the region given by pink circle in target has been sharply recovered by DC-RSN; DC-Hybrid and DC-CNN has partially recovered the structures, while ReGAN provided a smooth region without the structures.}
    \label{fig:calgary}
\end{figure*}

In this experiment, Calgary dataset was used to compare DC-RSN with the architectures proposed for complex-valued single-coil MRI reconstruction. The comparison of DC-RSN with other architectures is presented in Table \ref{tab:calgary}. From the table, it is seen that, deep cascade architectures are better than ReGAN. It also can be noticed that DC-Hybrid is better than DC-CNN. This is due to alternate CNN operating on k-space and image domain unlike DC-CNN, which operates only on the image domain. However, DC-RSN is significantly better than DC-Hybrid showing that RSN has effectively utilized k-space and image information. The comparison of architectures with an example image is depicted in Fig. \ref{fig:calgary}. The better structure recovery of complex structures in DC-RSN compared to other architectures can be noticed in the figure.

\begin{table}[]
\scriptsize
\centering
\caption{Quantitative comparison of single-coil complex valued MRI reconstruction architectures}
\label{tab:calgary}
\begin{tabular}{|l|l|l|l|l|}
\hline
\multirow{3}{*}{Method} & \multicolumn{4}{c|}{Calgary} \\ \cline{2-5} 
 & \multicolumn{2}{c|}{4x} & \multicolumn{2}{c|}{5x} \\ \cline{2-5} 
 & \multicolumn{1}{c|}{PSNR} & \multicolumn{1}{c|}{SSIM} & \multicolumn{1}{c|}{PSNR} & \multicolumn{1}{c|}{SSIM} \\ \hline
US & 26.91 $\pm$ 0.9 & 0.740 $\pm$ 0.0 & 26.49 $\pm$ 0.9 & 0.727 $\pm$ 0.0 \\ \hline
ReGAN\cite{cyclic_loss} & 33.71 $\pm$ 0.9 & 0.921 $\pm$ 0.0 & 33.1 $\pm$ 0.94 & 0.91 $\pm$ 0.0 \\ \hline
DC-CNN\cite{dc_cnn}& 36.66 $\pm$ 0.9 & 0.952 $\pm$ 0.0 & 35.22 $\pm$ 0.9 & 0.937 $\pm$ 0.0 \\ \hline
DC-Hybrid\cite{hybrid} & 36.85 $\pm$ 0.9 & 0.954 $\pm$ 0.0 & 35.52 $\pm$ 0.9 & 0.941 $\pm$ 0.0 \\ \hline
DC-RSN(Ours) & \textbf{37.85 $\pm$ 1.0} & \textbf{0.962 $\pm$ 0.0} & \textbf{36.04 $\pm$ 1.0} & \textbf{0.948 $\pm$ 0.0} \\ \hline
\end{tabular}
\end{table}

\begin{table*}[]
\scriptsize
\centering
\caption{Quantitative comparison of multi-coil MRI reconstruction architectures}
\label{tab:multi_coil_knee}
\begin{tabular}{|l|l|l|l|l|l|l|l|l|}
\hline
\multirow{3}{*}{Method} & \multicolumn{4}{c|}{Coronal PD} & \multicolumn{4}{c|}{Coronal fat-saturated PD} \\ \cline{2-9} 
 & \multicolumn{2}{c|}{4x} & \multicolumn{2}{c|}{5x} & \multicolumn{2}{c|}{4x} & \multicolumn{2}{c|}{5x} \\ \cline{2-9} 
 & \multicolumn{1}{c|}{PSNR} & \multicolumn{1}{c|}{SSIM} & \multicolumn{1}{c|}{PSNR} & \multicolumn{1}{c|}{SSIM} & \multicolumn{1}{c|}{PSNR} & \multicolumn{1}{c|}{SSIM} & \multicolumn{1}{c|}{PSNR} & \multicolumn{1}{c|}{SSIM} \\ \hline
US & 31.42 $\pm$ 3.92 & 0.884 $\pm$ 0.06 & 29.26 $\pm$ 3.91 & 0.8465 $\pm$ 0.07 & 33.44 $\pm$ 2.95 & 0.8535 $\pm$ 0.07 & 31.67 $\pm$ 2.87 & 0.8109 $\pm$ 0.09 \\ \hline
VN \cite{variational} & 39.8 $\pm$ 3.77 & 0.9595 $\pm$ 0.02 & 34.15 $\pm$ 3.35 & 0.9122 $\pm$ 0.04 & 37.26 $\pm$ 3.61 & 0.8875 $\pm$ 0.07 & 34.45 $\pm$ 3.01 & 0.8385 $\pm$ 0.09 \\ \hline
VS-Net \cite{vs_net} & 39.87 $\pm$ 3.78 & 0.9604 $\pm$ 0.02 & 33.95 $\pm$ 3.34 & 0.9114 $\pm$ 0.04 & 37.32 $\pm$ 3.63 & 0.8883 $\pm$ 0.07 & 34.34 $\pm$ 2.93 & 0.8389 $\pm$ 0.08 \\ \hline
VS-RSN (Ours) & \textbf{40.45 $\pm$ 3.94} & \textbf{0.9636 $\pm$ 0.02} & \textbf{35.67 $\pm$ 3.51} & \textbf{0.9293 $\pm$ 0.03} & \textbf{37.43 $\pm$ 3.68} & \textbf{0.8914 $\pm$ 0.07} & \textbf{34.79 $\pm$ 3.26} & \textbf{0.8453 $\pm$ 0.09} \\ \hline
\multirow{3}{*}{} & \multicolumn{4}{c|}{Sagittal fat-saturated T2} & \multicolumn{4}{c|}{Sagittal PD} \\ \cline{2-9} 
 & \multicolumn{2}{c|}{4x} & \multicolumn{2}{c|}{5x} & \multicolumn{2}{c|}{4x} & \multicolumn{2}{c|}{5x} \\ \cline{2-9} 
 & \multicolumn{1}{c|}{PSNR} & \multicolumn{1}{c|}{SSIM} & \multicolumn{1}{c|}{PSNR} & \multicolumn{1}{c|}{SSIM} & \multicolumn{1}{c|}{PSNR} & \multicolumn{1}{c|}{SSIM} & \multicolumn{1}{c|}{PSNR} & \multicolumn{1}{c|}{SSIM} \\ \hline
US & 38.08 $\pm$ 3.81 & 0.936 $\pm$ 0.04 & 37.25 $\pm$ 3.82 & 0.9271 $\pm$ 0.05 & 39.27 $\pm$ 2.98 & 0.9643 $\pm$ 0.01 & 38.8 $\pm$ 2.99 & 0.9606 $\pm$ 0.01 \\ \hline
VN \cite{variational} & 41.82 $\pm$ 4.09 & 0.9498 $\pm$ 0.04 & 40.54 $\pm$ 4.06 & 0.9399 $\pm$ 0.05 &  43.86 $\pm$ 2.63 & 0.9788 $\pm$ 0.00 & 41.89 $\pm$ 2.80 & 0.9724 $\pm$ 0.01 \\ \hline
VS-Net \cite{vs_net} & 41.84 $\pm$ 4.1 & 0.9491 $\pm$ 0.04 & 40.63 $\pm$ 4.05 & 0.94 $\pm$ 0.05 & 44.25 $\pm$ 2.58 & 0.9793 $\pm$ 0.00 & 42.2 $\pm$ 2.73 & 0.9731 $\pm$ 0.01 \\ \hline
VS-RSN & \textbf{41.98 $\pm$ 4.18} & \textbf{0.951 $\pm$ 0.04} & \textbf{40.88 $\pm$ 4.12} & \textbf{0.9419 $\pm$ 0.05} & \textbf{44.3 $\pm$ 2.60} & \textbf{0.9805 $\pm$ 0.00} & \textbf{42.5 $\pm$ 2.72} & \textbf{0.975 $\pm$ 0.00} \\ \hline
\end{tabular}
\end{table*}

\begin{figure*}
    \centering
    \includegraphics[width=0.9\linewidth]{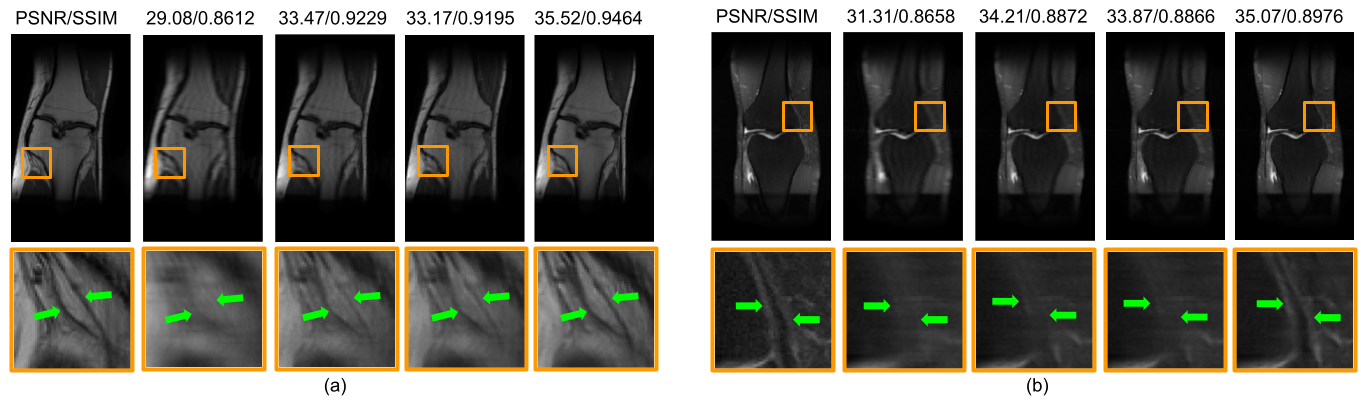}
    \caption{(a) Qualitative comparison of multi-coil MRI reconstruction architectures for coronal PD. From left to right (Target, ZF, VN, VS-Net, and VS-RSN). The structures denoted by green arrows in target have been faithfully recovered by VS-RSN compared to VN and VS-Net. VS-RSN has thoroughly delineated the marked structures from its background, while VN and VS-Net provided a blurry output. (b) Qualitative comparison of multi-coil MRI reconstruction architectures for coronal fat-saturated PD. From left to right (Target, ZF, VN, VS-Net, and VS-RSN). The structures denoted by green arrows in target have been sharply recovered by VS-RSN. VN and VS-Net have missed both the structures resulting in a lack of boundary.}
    \label{fig:multi_coil_knee}
\end{figure*}

\subsection{Multi-coil MRI reconstruction}
In this experiment, VS-RSN is compared with the state-of-the-art architectures in multi-coil acquisition for different protocols in knee dataset. PSNR and SSIM metric comparison is presented in Table \ref{tab:multi_coil_knee}. It is observed that VN and VS-Net show similar performance, while VS-RSN fairs better than both VN and VS-Net for different acceleration factors and protocols. This shows the successful incorporation of sophisticated RSN block as a denoiser in VS-Net, thereby translating RSN for a multi-coil setting. The qualitative comparison of the reconstruction methods for coronal PD and coronal fat-saturated PD acquisition protocols is presented in Fig. \ref{fig:multi_coil_knee}, which shows that VS-RSN is able to delineate complex structures in comparison to VS-Net and VN where the structures look fuzzy. Quantitative comparison of axial protocol can be found in Table A1 of the supplementary material. 

\subsection{Assistance to MRI reconstruction}
\subsubsection{Gradient assistance}
\begin{figure}
    \centering
    \includegraphics[width=0.9\linewidth]{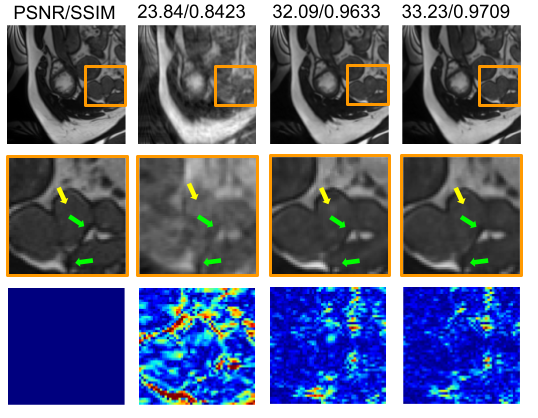}
    \caption{Qualitative comparison of DC-RSN and DC-RSN-GOLF. From left to right (Target, ZF, DC-RSN, and DC-RSN-GOLF). The green arrows in target indicate challenging structures. These structures have been recovered by DC-RSN-GOLF, while missed by DC-RSN. The yellow arrow shows a region where DC-RSN-GOLF restores the degradation caused by undersampling, while it is not handled by DC-RSN.}
    \label{fig:gradient_assistance}
\end{figure}

In this experiment, the validation of the effect of GOLF assistance to DC-RSN and VS-RSN were carried out. The quantitative comparison of the architecture with and without GOLF assistance is presented in Table \ref{tab:gradient_assistance}. It is clearly seen that GOLF assistance provides a substantial improvement in evaluation metrics across acceleration factors and datasets. The respective qualitative comparison is shown with an example in Fig. \ref{fig:gradient_assistance}. It is noticed that the architecture with GOLF enhances subtle structures and appreciably recovers grainy regions as compared to the one obtained without GOLF. 

\begin{table}[]
\scriptsize
\centering
\caption{Quantitative comparison of DC-RSN, VS-RSN with and without GOLF assistance}
\label{tab:gradient_assistance}
\begin{tabular}{|l|l|l|l|l|l|}
\hline
\multirow{2}{*}{Dataset} & \multirow{2}{*}{Method} & \multicolumn{2}{c|}{4x} & \multicolumn{2}{c|}{5x} \\ \cline{3-6} 
 &  & \multicolumn{1}{c|}{PSNR} & \multicolumn{1}{c|}{SSIM} & \multicolumn{1}{c|}{PSNR} & \multicolumn{1}{c|}{SSIM} \\ \hline
\multirow{2}{*}{Cardiac} & DC-RSN & 33.61 & 0.9322 & 32.65 & 0.92 \\ \cline{2-6} 
 & DC-RSN-GOLF & \textbf{33.68} & \textbf{0.933} & \textbf{32.75} & \textbf{0.9214} \\ \hline
\multirow{2}{*}{Kirby} & DC-RSN & 36.83 & 0.9707 & 35.22 & 0.9609 \\ \cline{2-6} 
 & DC-RSN-GOLF & \textbf{36.92} & \textbf{0.9723} & \textbf{35.27} & \textbf{0.9618} \\ \hline
\multirow{2}{*}{Coronal-pd} & VS-RSN & 40.45 & 0.9636 & 35.67 & 0.9293 \\ \cline{2-6} 
 & VS-RSN-GOLF & \textbf{40.52} & \textbf{0.9645} & \textbf{35.83} & \textbf{0.9316} \\ \hline
\end{tabular}
\end{table}

\subsubsection{T1 assistance}
\begin{figure*}[]
    \centering
    \includegraphics[width=0.9\linewidth]{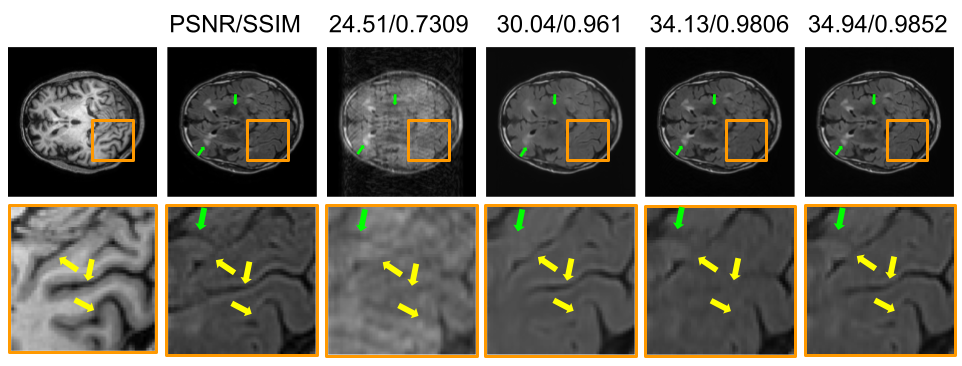}
    \caption{ Qualitative comparison of DenseUNet-T1, DC-RSN, and DC-RSN-T1. From left to right (T1 FS, T2 FS, T2 ZF, DenseUNet-T1, DC-RSN, and DC-RSN-T1). The structures denoted by yellow arrows in T2 FS has not been recovered by DC-RSN, while DenseUNet-T1 and DC-RSN-T1 have recovered those structures through FS T1 assistance. Green arrows indicate regions in DC-RSN-T1 and DC-RSN which are closer to FS T2 than DenseUNet-T1.}
    \label{fig:t1assistance}
\end{figure*}

This experiment was designed to understand the contribution of T1 assistance. The quantitative comparison of DenseUnet-T1 (Xiang et al., 2019), DC-RSN, and DC-RSN-T1 is presented in Table \ref{tab:t1assistance}. From table, it is observed that DC-RSN-T1 is better than both DC-RSN and DenseUNet-T1 in terms of PSNR and SSIM for different acceleration factors. Further, on closer inspection of Fig. \ref{fig:t1assistance}, it is noticed that DC-RSN-T1 and DenseUNet-T1 have reconstructed a structure which is completely missed by DC-RSN. This is due to FS T1 assistance in DenseUNet-T1 and DC-RSN-T1. 

\begin{table}[]
\scriptsize
\centering
\caption{Quantitative comparison of different combinations of T1 and GOLF with DC-RSN}
\label{tab:t1assistance}
\begin{tabular}{|l|l|l|l|l|}
\hline
\multirow{2}{*}{Method} & \multicolumn{2}{c|}{4x} & \multicolumn{2}{c|}{5x} \\ \cline{2-5} 
 & \multicolumn{1}{c|}{PSNR} & \multicolumn{1}{c|}{SSIM} & \multicolumn{1}{c|}{PSNR} & \multicolumn{1}{c|}{SSIM} \\ \hline
US & 28.4 & 0.6422 & 26.99 & 0.6095 \\ \hline
DenseUNet-T1 \cite{t1_assistance} & 34.2 & 0.9428 & 32.88 & 0.9315 \\ \hline
DC-RSN & 37.76 & 0.9731 & 37.05 & 0.9643 \\ \hline
DC-RSN-GOLF & 38 & 0.9742 & 37.44 & 0.9675 \\ \hline
DC-RSN-T1 & 38.34 & 0.976 & 37.66 & 0.9705 \\ \hline
DC-RSN-T1-GOLF & \textbf{38.6} & \textbf{0.9774} & \textbf{38.08} & \textbf{0.9722} \\ \hline
\end{tabular}
\end{table}

\begin{figure*}
    \centering
    \includegraphics[width=0.9\linewidth]{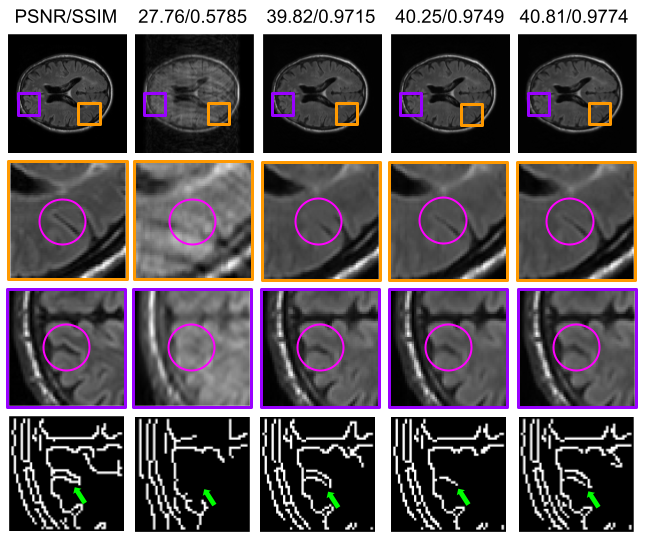}
    \caption{Qualitative comparison of different combinations of T1 and GOLF with DC-RSN. From left to right (Target, ZF, DC-RSN-GOLF, DC-RSN-T1, and DC-RSN-T1-GOLF). The structure indicated by pink circle in the region denoted by orange in the target has been recovered by DC-RSN-T1 and DC-RSN-T1-GOLF, while DC-RSN-GOLF has failed to recover the entire structure. The edge map of the structure indicated by pink circle in the region denoted by purple in the target image is closer to DC-RSN-GOLF and DC-RSN-T1-GOLF compared to DC-RSN-T1. The green arrow in the edge map indicates the structure of interest. }
    \label{fig:gradient_t1_assistance}
\end{figure*}

\subsubsection{Gradient based T1 assistance}
In this experiment, the performance of DC-RSN-GOLF, DC-RSN-T1 and DC-RSN-T1-GOLF were compared. The data of the quantitative and the qualitative comparison are provided in Table \ref{tab:t1assistance} and in Fig. 10, respectively. From the table the following are observed: 1) DC-RSN-T1 is better than DC-RSN-GOLF as T1 assistance has provided missing structures while GOLF assistance has only enhanced the existing structures; 2) DC-RSN-T1-GOLF is better than DC-RSN-GOLF as GOLF obtained using DC-RSN-T1 is better than the one obtained using DC-RSN; and, 3) DC-RSN-T1-GOLF is better than DC-RSN-T1. This is because GOLF containing the structural information of DC-RSN-T1 is explicitly provided to DC-RSN for final prediction. From the figure it is seen that DC-RSN-T1-GOLF and DC-RSN-T1 has reconstructed structures missed by DC-RSN-GOLF. In addition, DC-RSN-T1-GOLF and DC-RSN-GOLF has recovered subtle structures compared to DC-RSN-T1. This structure improvement can be better appreciated through edge maps of the region of interest; moreover, restoring the image gradient is the primary motivation behind GOLF assistance.

\subsection{PRN for MRI reconstruction}
In this experiment, PRN is validated by using it to refine the reconstructions of the proposed networks DC-RSN, VS-RSN, and DC-RSN-T1-GOLF. The comparison of networks with and without PRN block is provided in Table VI. From the table, it is observed that addition of PRN has improved VIF across acceleration factors, datasets, and networks. Additionally, improvement in VIF has reduced PSNR and SSIM to some extent which is expected as PRN is trained with higher weightage to adversarial term (Blau and Michaeli, 2018). Sample reconstructions of networks using PRN can be found in Fig. A3, A4, and A5  of the supplementary material. The quantitative metrics of models developed for better perceptual quality including DAGAN, ReGAN, and VN is also added in Table VI. It is observed that VIF for these models are significantly lower than our proposed models. 

\begin{table*}[]
\scriptsize
\centering
\caption{Quantitative comparison of DC-RSN, VS-RSN, DC-RSN-T1-GOLF with and without PRN.}
\label{tab:perceptual_refinement}
\begin{tabular}{|l|l|l|l|l|l|l|l|}
\hline
\multirow{2}{*}{Dataset} & \multirow{2}{*}{Method} & \multicolumn{3}{c|}{4x} & \multicolumn{3}{c|}{5x} \\ \cline{3-8} 
 &  & \multicolumn{1}{c|}{PSNR} & \multicolumn{1}{c|}{SSIM} & \multicolumn{1}{c|}{VIF} & \multicolumn{1}{c|}{PSNR} & \multicolumn{1}{c|}{SSIM} & \multicolumn{1}{c|}{VIF} \\ \hline
\multirow{2}{*}{Cardiac} & DC-RSN & 33.36 $\pm$ 3.55 & 0.9279 $\pm$ 0.04 & 0.928 $\pm$ 0.05 & 32.56 $\pm$ 3.68 & 0.9188 $\pm$ 0.04 & 0.92 $\pm$ 0.05 \\ \cline{2-8} 
 & DC-RSN-PRN & 33.39 $\pm$ 3.55 & 0.9295 $\pm$ 0.03 & 0.959 $\pm$ 0.04 & 32.36 $\pm$ 3.59 & 0.9167 $\pm$ 0.04 & 0.951 $\pm$ 0.05 \\ \hline
\multirow{2}{*}{Calgary} & DC-RSN & 37.85 $\pm$ 1.08 & 0.9621 $\pm$ 0.00 & 0.946 $\pm$ 0.01 & 36.04 $\pm$ 1.00 & 0.9483 $\pm$ 0.01 & 0.939 $\pm$ 0.01 \\ \cline{2-8} 
 & DC-RSN-PRN & 37.75 $\pm$ 1.05 & 0.9612 $\pm$ 0.00 & 0.982 $\pm$ 0.01 & 35.84 $\pm$ 0.96 & 0.9466 $\pm$ 0.01 & 0.992 $\pm$ 0.02 \\ \hline
\multirow{2}{*}{Coronal PD} & VS-RSN & 40.45 $\pm$ 3.94 & 0.9636 $\pm$ 0.02 & 0.951 $\pm$ 0.04 & 35.67 $\pm$ 3.51 & 0.9293 $\pm$ 0.03 & 0.851 $\pm$ 0.07 \\ \cline{2-8} 
 & VS-RSN-PRN & 40.26 $\pm$ 3.62 & 0.9647 $\pm$ 0.02 & 0.978 $\pm$ 0.04 & 34.42 $\pm$ 3.33 & 0.9166 $\pm$ 0.04 & 0.879 $\pm$ 0.11 \\ \hline
\multirow{2}{*}{MRBrains} & DC-RSN-T1-GOLF & 38.6 $\pm$ 0.27 & 0.9774 $\pm$ 0.00 & 0.962 $\pm$ 0.04 & 38.08 $\pm$ 0.49 & 0.9722 $\pm$ 0.00 & 0.979 $\pm$ 0.05 \\ \cline{2-8} 
 & DC-RSN-T1-GOLF-PRN & 38.2 $\pm$ 0.27 & 0.9755 $\pm$ 0.00 & 0.978 $\pm$ 0.03 & 38.16 $\pm$ 0.51 & 0.969 $\pm$ 0.00 & 0.979 $\pm$ 0.04 \\ \hline
\end{tabular}
\end{table*}

\subsection{Ablative study}

\begin{table*}[]
\scriptsize
\centering
\caption{Quantitative comparison of ablative study on RSN}
\label{tab:ablative-study}
\begin{tabular}{|l|l|l|l|l|l|l|l|}
\hline
\multirow{2}{*}{} & \multicolumn{1}{c|}{\multirow{2}{*}{Method}} & \multicolumn{2}{c|}{2x} & \multicolumn{2}{c|}{2.5x} & \multicolumn{2}{c|}{3.3x} \\ \cline{3-8} 
 & \multicolumn{1}{c|}{} & \multicolumn{1}{c|}{PSNR} & \multicolumn{1}{c|}{SSIM} & \multicolumn{1}{c|}{PSNR} & \multicolumn{1}{c|}{SSIM} & \multicolumn{1}{c|}{PSNR} & \multicolumn{1}{c|}{SSIM} \\ \hline
\multirow{10}{*}{\begin{tabular}[c]{@{}l@{}}Lower \\ acceleration \\ factors\end{tabular}} & US & 29.63 $\pm$ 3.17 & 0.8435 $\pm$ 0.05 & 26.71 $\pm$ 3.14 & 0.7582 $\pm$ 0.07 & 26.95 $\pm$ 3.12 & 0.7906 $\pm$ 0.06 \\ \cline{2-8} 
 & KK $\rightarrow$ IFT & 32.55 $\pm$ 2.90 & 0.8919 $\pm$ 0.03 & 28.75 $\pm$ 2.89 & 0.8133 $\pm$ 0.05 & 29.77 $\pm$ 3.03 & 0.8755 $\pm$ 0.04 \\ \cline{2-8} 
 & II-CNN & 32.73 $\pm$ 2.94 & 0.9146 $\pm$ 0.03 & 29.55 $\pm$ 2.85 & 0.8559 $\pm$ 0.04 & 29.51 $\pm$ 2.88 & 0.8563 $\pm$ 0.05 \\ \cline{2-8} 
 & KI & 33.72 $\pm$ 3.09 & 0.9278 $\pm$ 0.03 & 29.98 $\pm$ 2.89 & 0.8638 $\pm$ 0.04 & 31.59 $\pm$ 3.15 & 0.908 $\pm$ 0.04 \\ \cline{2-8} 
 & II & 31.31 $\pm$ 3.34 & 0.928 $\pm$ 0.03 & 28.87 $\pm$ 3.15 & 0.8853 $\pm$ 0.04 & 28.49 $\pm$ 3.36 & 0.8825 $\pm$ 0.05 \\ \cline{2-8} 
 & II $\rightarrow$ FT $ \rightarrow$ KI & 31.2 $\pm$ 3.503 & 0.9228 $\pm$ 0.04 & 28.76 $\pm$ 3.25 & 0.8756 $\pm$ 0.05 & 28.4 $\pm$ 3.39 & 0.8786 $\pm$ 0.05 \\ \cline{2-8} 
 & KI $\rightarrow$ II & 31.73 $\pm$ 3.15 & 0.9265 $\pm$ 0.03 & 28.72 $\pm$ 2.87 & 0.8647 $\pm$ 0.04 & 29.56 $\pm$ 3.393 & 0.9053 $\pm$ 0.04 \\ \cline{2-8} 
 & Mean ( KI $\vert$ II ) & 33.61 $\pm$ 2.92 & 0.938 $\pm$ 0.02 & 30.53 $\pm$ 2.77 & 0.8905 $\pm$ 0.04 & 30.95 $\pm$ 2.93 & 0.9088 $\pm$ 0.04 \\ \cline{2-8} 
 & Fu ( KI $\vert$ II ) & 34.39 $\pm$ 2.84 & 0.9417 $\pm$ 0.02 & 31.13 $\pm$ 2.76 & 0.8958 $\pm$ 0.03 & 32.61 $\pm$ 3.07 & 0.9235 $\pm$ 0.04 \\ \cline{2-8} 
 & Fu ( KI $\vert$ II $\vert$ US) & 34.72 $\pm$ 2.89 & 0.9427 $\pm$ 0.02 & 31.29 $\pm$ 2.81 & 0.8981 $\pm$ 0.03 & 32.72 $\pm$ 3.0 & 0.9249 $\pm$ 0.03 \\ \hline
\multirow{12}{*}{\begin{tabular}[c]{@{}l@{}}Higher\\ acceleration \\ factors\end{tabular}} & \multicolumn{1}{c|}{\multirow{2}{*}{Model}} & \multicolumn{2}{c|}{4x} & \multicolumn{2}{c|}{5x} & \multicolumn{2}{c|}{8x} \\ \cline{3-8} 
 & \multicolumn{1}{c|}{} & \multicolumn{1}{c|}{PSNR} & \multicolumn{1}{c|}{SSIM} & \multicolumn{1}{c|}{PSNR} & \multicolumn{1}{c|}{SSIM} & \multicolumn{1}{c|}{PSNR} & \multicolumn{1}{c|}{SSIM} \\ \cline{2-8} 
 & US & 24.27 $\pm$ 3.10 & 0.6996 $\pm$ 0.08 & 23.82 $\pm$ 3.11 & 0.6742 $\pm$ 0.08 & 22.83 $\pm$ 3.11 & 0.6344 $\pm$ 0.09 \\ \cline{2-8} 
 & KK $\rightarrow$ IFT & 26.76 $\pm$ 2.89 & 0.7622 $\pm$ 0.06 & 25.88 $\pm$ 2.86 & 0.73 $\pm$ 0.06 & 24.53 $\pm$ 2.83 & 0.6772 $\pm$ 0.07 \\ \cline{2-8} 
 & II-CNN & 26.86 $\pm$ 2.87 & 0.784 $\pm$ 0.06 & 26.4 $\pm$ 2.90 & 0.7642 $\pm$ 0.06 & 25.19 $\pm$ 2.97 & 0.724 $\pm$ 0.07 \\ \cline{2-8} 
 & KI & 27.85 $\pm$ 2.87 & 0.8092 $\pm$ 0.06 & 26.91 $\pm$ 2.95 & 0.7829 $\pm$ 0.06 & 25.51 $\pm$ 2.96 & 0.7299 $\pm$ 0.08 \\ \cline{2-8} 
 & II & 26.98 $\pm$ 3.14 & 0.8409 $\pm$ 0.06 & 26.89 $\pm$ 3.04 & 0.8327 $\pm$ 0.06 & 25.3 $\pm$ 2.93 & 0.7796 $\pm$ 0.07 \\ \cline{2-8} 
 & II $\rightarrow$ FT $\rightarrow$ KI & 26.99 $\pm$ 3.15 & 0.8334 $\pm$ 0.06 & 26.78 $\pm$ 3.07 & 0.8245 $\pm$ 0.06 & 25.17 $\pm$ 2.92 & 0.7697 $\pm$ 0.08 \\ \cline{2-8} 
 & KI $\rightarrow$ II & 26.64 $\pm$ 2.83 & 0.8108 $\pm$ 0.06 & 25.98 $\pm$ 2.81 & 0.7851 $\pm$ 0.06 & 24.67 $\pm$ 2.70 & 0.7365 $\pm$ 0.08 \\ \cline{2-8} 
 & Mean ( KI $\vert$ II ) & 28.56 $\pm$ 2.72 & 0.8459 $\pm$ 0.05 & 27.94 $\pm$ 2.81 & 0.8288 $\pm$ 0.06 & 26.28 $\pm$ 2.82 & 0.7764 $\pm$ 0.07 \\ \cline{2-8} 
 & Fu ( KI $\vert$ II ) & 28.68 $\pm$ 2.63 & 0.8525 $\pm$ 0.05 & 27.8 $\pm$ 2.82 & 0.8376 $\pm$ 0.06 & 26.2 $\pm$ 2.75 & 0.7861 $\pm$ 0.07 \\ \cline{2-8} 
 & Fu ( KI $\vert$ II $\vert$ US) & 28.78 $\pm$ 2.62 & 0.855 $\pm$ 0.05 & 27.8 $\pm$ 2.73 & 0.8367 $\pm$ 0.05 & 26.19 $\pm$ 2.73 & 0.7856 $\pm$ 0.07 \\ \hline
\end{tabular}
\end{table*}

\begin{figure*}
    \centering
    \includegraphics[width=0.9\linewidth]{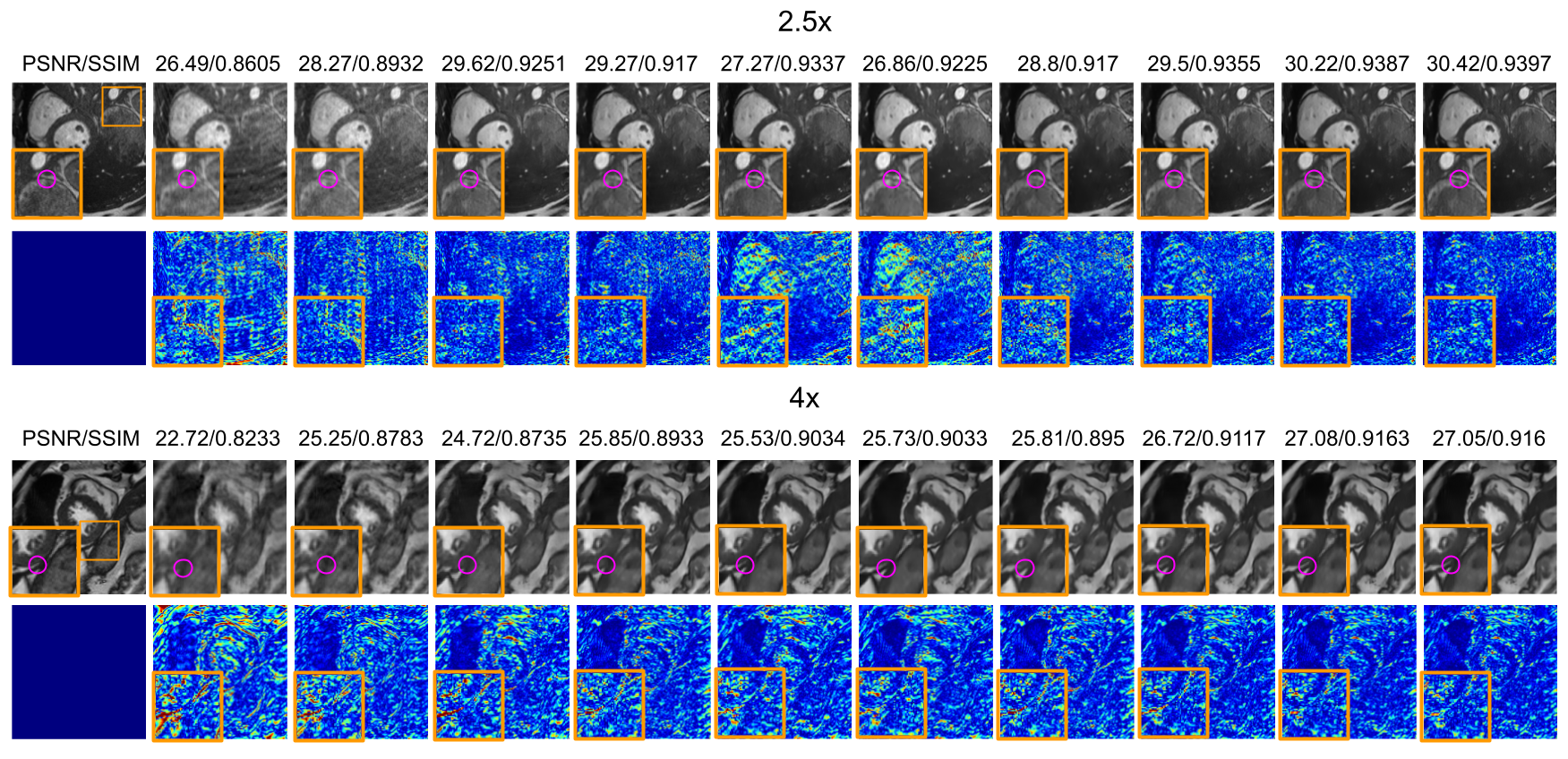}
    \caption{Qualitative results of ablative study. From left to right (FS, US, KK $\rightarrow$ IFT, II-CNN, KI, II, II $\rightarrow$ FT $\rightarrow$ KI, KI $\rightarrow$ II, Mean ( KI $\vert$ II ), Fu ( KI $\vert$ II ) , Fu ( KI $\vert$ II $\vert$ US)). }
    \label{fig:ablative_study}
\end{figure*}
An ablative study was conducted to understand the role of individual networks (KI, II, Fu) in RSN. The quantitative comparison of the combinations of different networks for various acceleration factors is presented in Table VII. KI (dAUTOMAP) is better than KK $\rightarrow$ IFT (k-space CNN followed by IFT) in both the metrics for all acceleration factors. This shows that KI is a better choice to move from k-space to image due to its specifically designed convolution layers (domain transform layers) instead of normal 2D convolutions in KK $\rightarrow$ IFT. II (image space UNet) is significantly better than II-CNN (image space CNN) in SSIM for every acceleration factor and competitively closer in PSNR. The superior SSIM can be attributed to encoder-decoder multi scale network design of UNet, making it an ideal choice to operate on image domain for structure recovery. Between KI and II, it is observed that KI has higher PSNR (lower reconstruction error) while II has higher SSIM (better structure recovery). This shows that effective combination of KI and II can produce both better PSNR and SSIM.

The proposed Fu (KI | II) network presented here provided significantly better PSNR and SSIM compared to Mean (KI | II) for lower acceleration factors, while for higher acceleration factors, Fu (KI | II) provided better SSIM and competitively closer PSNR. Similarly, the network Fu (KI | II | US) provided better metrics than Fu (KI | II) for lower acceleration factors, but for higher acceleration factors both showed similar results. These observations show that CNN acts as an effective fusion network. The disparity in the network's performance for lower and higher acceleration factors is due to varying sparsity of frequency components in k-space which impacts the performance of KI for higher acceleration factors. The qualitative comparison of different combinations of networks mentioned in this section is provided in Fig. 11. The following are observed in the figure: 1) the networks starting with k-space domain ( KI, KI $\rightarrow$ II) provided lower residue; 2) the networks starting with image domain ( II, II $\rightarrow$ FT $\rightarrow$ KI ) provided better structure recovery; 3) networks simultaneously operating on both domains ( Mean ( KI $\vert$ II ), Fu ( KI $\vert$ II ) , Fu ( KI $\vert$ II $\vert$ US))) provided lower residue and better structure recovery; and 4) RSN ( Fu (KI | II),  Fu (KI | II | US) ) provided enhanced structures compared to Mean (KI | II).

\subsection{RSN and different US masks}
Evaluation of RSN for other standard US masks (Gaussian, Radial, and Spiral) was carried out by comparing DC-CNN and DC-RSN in a single cascade mode for MRBrains T1 dataset. The US data was prepared using 5x US masks of ReGAN (Quan et al., 2018). The quantitative comparison of DC-CNN and DC-RSN for different US masks is depicted in Fig. 12. It is observed that DC-RSN outperforms DC-CNN for every US mask, demonstrating that RSN can be used across standard masks. 

\begin{figure*}
    \centering
    \includegraphics[width=0.9\linewidth]{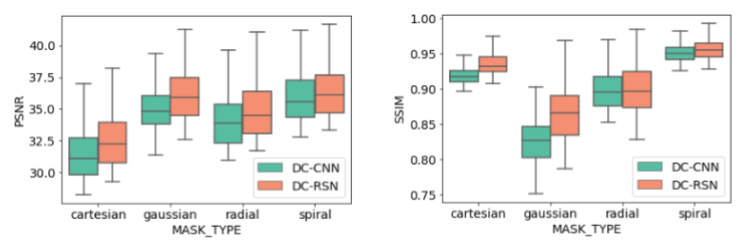}
    \caption{Quantitative comparison of DC-CNN and DC-RSN for different US masks.}
    \label{fig:box_plot}
\end{figure*}

\subsection{Misregistration environment}
In our experiments, it has been assumed that the sequences T1WI and T2WI have undergone registration, i.e., perfectly aligned. However, in real MRI scenarios, such an accurate registration is not always possible. Hence, in this experiment, the performance of DC-RSN-T1 for a fixed T2WI with a randomly shifted T1WI for a maximum of 2 pixels is investigated as followed in DISN (Sun et al., 2019b). It is observed that SSIM of DC-RSN-T1 consistently dropped for different possible shifts, and it was also noticed that T1 assistance did not aid in the recovery of missing structures. To make DC-RSN-T1 robust to these random shifts, the model is trained with fixed T2WI and randomly shifted T1WI for upto 2 pixels in both x and y directions. It is noted that the DC-RSN-T1 trained for these random shifts are robust to those shifts, and it also helped in recovering structures that were degraded in US T2WI. The respective quantitative and qualitative comparison are presented in Fig. 13. 

\begin{figure*}
    \centering
    \includegraphics[width=0.9\linewidth]{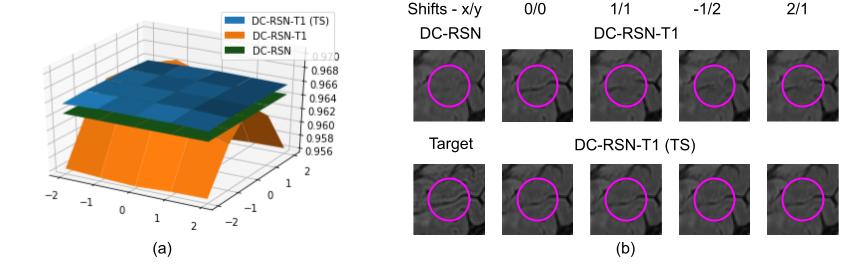}
    \caption{ Quantitative (a) and qualitative (b) comparison of model trained with perfectly registered pairs (DC-RSN-T1) and model trained with randomly shifted pairs (DC-RSN-T1 (TS)). The baseline is given by DC-RSN. SSIM for DC-RSN-T1 (TS) for various shifts is almost constant, while SSIM for DC-RSN-T1 drops significantly with increase in shifts in either direction. The structure recovered by DC-RSN-T1 (TS) for different shifts looks similar to the one without any shift, while the structure recovered by DC-RSN-T1 is severely affected by different shifts. }
    \label{fig:misregistration}
\end{figure*}

\subsection{Comparison with fastMRI}
DC-RSN and VS-RSN were evaluated with the single and multi-coil knee data of fastMRI, respectively. Results are reported at the public leaderboard , with team name HTIC and model name Cascade Hybrid. In single-coil track, DC-RSN with five cascades provided PSNR of 33.81 dB and SSIM of 0.768. In multi-coil track, VS-RSN with five cascades provided PSNR of 38.59 dB, and SSIM of 0.923 for 4x acceleration; and PSNR of 35.32 dB, and SSIM of 0.878 for 8x acceleration. 

\section{Conclusion}
\label{sec:conclusion}
In this work, we introduced RSN, a base network specifically designed to handle both the k-space and the image input for MRI reconstruction. Using RSN, we proposed DC-RSN and VS-RSN for high quality image reconstruction from US k-space of single- and multi-coil acquisitions. We enhanced the structure recovery of DC-RSN for T2WI reconstruction through GOLF based T1 assistance. We also presented PRN to improve the perceptual quality of reconstructions with respect to radiologist's opinion. We conducted an extensive study across datasets and acceleration factors and found the following: 1) DC-RSN and VS-RSN are better than respective state-of-the-art methods; 2) GOLF based T1 assistance provides more faithful reconstruction; and 3) PRN addition increases VIF, a metric highly correlated with the radiologist's opinion on image quality. 

The reconstructions in our work are evaluated using VIF metric. It will be interesting to conduct a study and evaluate the reconstructions with radiologists to better understand the correlation between VIF and scores from radiologists. Furthermore, the methods need to be evaluated for faithful reconstruction in pathology cases. In the case of network design, the feature fusion operations can be improved through attention mechanisms. 

{\small
\bibliographystyle{ieee_fullname}
\bibliography{cvpr_review}
}

\end{document}